\documentstyle[12pt,a4,epsfig]{article}
\def\dfrac#1#2{{\displaystyle#1\over\displaystyle#2}}
\begin{document}

\title
{Galactic Synchrotron Foreground and the CMB Polarization Measurements}

\author{M.V.Sazhin$^1$, G.Sironi$^2$, O.S.Khovanskaya$^1$
\footnote{E-mail: sazhin@sai.msu.su, ~giorgio.sironi@mib.infn.it,
~khovansk@sai.msu.ru}}

\maketitle
\begin{center}
\small{ $^1$ Sternberg Astronomical Institute, Universitetsky pr. 13, 119899
Moscow Russia

$^2$ Dipartimento di Fisica G. Occhialini, Universita` degli Studi di
Milano-Bicocca, Piazza della Scienza 3, I20126 Milano Italy}

\end{center}

\begin{center}
{\bf Abstract}
\end{center}
The polarization of the Cosmic Microwave Background (CMB)is a powerful
observational tool at hand for modern cosmology. It allows to break the
degeneracy of fundamental cosmological parameters one cannot obtain using only
anisotropy data and provides new insight into conditions existing in the very
early Universe. Many experiments are now in progress whose aim is detecting
anisotropy and polarization of the CMB. Measurements of the CMB polarization
are however hampered by the presence of polarized foregrounds, above all the
synchrotron emission of our Galaxy, whose importance increases as frequency
decreases and dominates the polarized diffuse radiation at frequencies below
$\simeq 50$ GHz. In the past the separation of CMB and synchrotron was made
combining observations of the same area of sky made at different frequencies.
In this paper we show that the statistical properties of the polarized
components of the synchrotron and dust foregrounds are different from the
statistical properties of the polarized component of the CMB, therefore one
can build a statistical estimator which allows to extract the polarized
component of the CMB from single frequency data also when the polarized CMB
signal is just a fraction of the total polarized signal. This estimator
improves the signal/noise ratio for the polarized component of the CMB and
reduces from $\simeq$50 GHz to $\simeq$20 GHz the frequency above which the
polarized component of the CMB can be extracted from single frequency maps of
the diffuse radiation. \vskip2cm

\section{Introduction}

This year is a decade since the first detection of the anisotropy of the
Cosmic Microwave Background at large angular scales ($\ge 10^0$)
\cite{str92}, \cite{smo92}. Today the CMB anisotropy (CMBA) has been detected
also at intermediate ($\sim (1^0 - 10^0)$) and small angular scales ($\le
1^0$), so the CMBA angular spectrum is now reasonably known down to the region
of the first and second Doppler peaks \cite{boo00}, \cite{boo01a},
\cite{boo01b}. Its shape gives information e.g. on the spectrum of the
primordial cosmological perturbations or can be used to test the inflation
theory but rises new questions to which CMBA cannot answers. Responses can on
the contrary be obtained looking at the CMB polarization (CMBP) produced by
Thomson scattering of CMB photons on the matter anisotropies at the
recombination epoch. In particular one can hope to use CMBP to disentangle the
effects of fundamental cosmological parameters like density of matter,
density of dark energy etc., effects anisotropy do not separate.This is among
the goals of space and ground based experiments like \cite{Map},
\cite{Planck1}, \cite{Planck2}, \cite{ami}, \cite{que}, \cite{gerva},
\cite{newboo} and is the main goal of SPOrt a polarization dedicated ASI/ESA
space mission on the International Space Station \cite{cor99}. The relevance
of the CMB polarization was remarked for the first time by M. Rees
\cite{ree65}. Since him many models of the expected features of the CMBP have
been published (see for instance \cite{saz95}, \cite{ng96}, \cite{mel97}).
They stimulated the search for CMBP, but in spite of many attempts so far no
CMB polarization has been detected. This observation is in fact extremely
difficult because the expected signal is at least an order of magnitude
smaller than the amplitude of the CMBA. Moreover foregrounds and their
inhomogeneities cover the polarized fraction of the CMB or mimic CMBP spots,
making the signal to noise (CMBP $/$ polarized foreground) ratio unfavorable.
In this paper we will concentrate on methods for improving this ratio and for
disentangling CMBP and polarized foregrounds.

In the microwave range the galactic foregrounds include:
\begin{itemize}
\item synchrotron radiation (strongly polarized),
\item free-free emission (polarization negligible),
\item dust radiation.
\end{itemize}

Because here we are interested in polarization, in the following  we will
neglect the free-free emission, whose expected level of polarization is
negligible. Moreover the polarized signals produced by dust, if present,
(e.g. \cite{set98}, \cite{fos01}), may be treated as an addition to the
synchrotron effects. In fact, as it will appears in the following, the
important quantities in our analysis are the statistical properties of the
foreground spatial distribution and, by good fortune, the spatial
distribution of the dust polarized emission is similar to that of the
synchrotron emission, since behind both radiation types there is the same
driving force, the galactic magnetic field which alignes dust grains and
guides radiating electrons.

Separation of foregrounds and CMBA was successfully solved when the
CMB anisotropy was discovered \cite{smo92}. When we go from CMBA to CMBP the
separation of foreground and background however is more demanding and the
problem of discriminating foreground inhomogneities from true CMB spots
severe. Approaches used in the past e.g. \cite{dod97}, \cite{teg99},
\cite{sto01}, \cite{teg96}, \cite{kog00}) were essentially based on the
differences between the frequency spectra of foregrounds and CMB, therefore
require multifrequency observations.

In this paper we suggest a different method which  takes advantage of the fact
that the measured values of the parameters we use to describes the
polarization of the diffuse radiation when measured at a given frequency in
different directions behave as stochastic variables. Because the mathematical
and statistical properties of these variables for synchrotron and CMB are
different, we suggest to use statistical methods for analyzing single
frequency maps of the diffuse radiation and disentangling their main
components, synchrotron and CMBP.
This method was proposed and briefly discussed in \cite{saz01}. Here we
present a more complete analysis.

\section{Polarization Parameters}
\subsection{The Stokes Parameters}
Convenient quantities commonly used to describe the polarization status of
radiation are the Stokes parameters (see for instance \cite{gin64},
\cite{gar66}, \cite{gin69}). Let's assume a monochromatic, plane, wave of
intensity I and amplitude $\propto \sqrt {I}$. In the {\it observer plane},
orthogonal to the direction of propagation of the electromagnetic wave
emitted by the electron, we can choose a pair of orthogonal axes ${\vec l}$
and ${\vec r}$. On that plane the amplitude vector of an unpolarized wave
moves in a random way. On the contrary it describes a figure, the {\it
polarization ellipse}, when the wave is polarized.

Projecting the wave amplitude on ${\vec l}$ and ${\vec r}$ we get two
orthogonal, linearly polarized, waves of intensity $I_l$ and $I_r$,($I = I_l
+ I_r$) whose amplitudes are $\propto \sqrt {I_l}$ and $\propto \sqrt {I_r}$
respectively. If the original wave of intensity $I$ is polarized, $I_l$ and
$I_r$ are correlated: let`s call $I_{12}$ and $I_{21}$ their correlation
products.

By definition the Stokes parameters are the four quantities: $I=I_l +I_r$,
$Q=I_l -I_r$, $U=I_{12} + I_{21}$, and $V= i(I_{21} - I_{12})$.

$Q$ and $U$ describe the linear polarization, $V$ the circular polarization
and $I$ the total intensity.

\noindent The ratio

\begin{equation}
\tan 2 \chi = \dfrac{U}{Q}
\end{equation}

\noindent gives the angle $\chi$ between the vector $ \vec l$ and the main
axis of the polarization ellipse ($0 < \chi < \pi$).

Rotating the $(l,r)$ coordinate system by an angle $\phi$ we get a
new coordinate system $(\hat l , \hat r)$ in which the Stokes parameters
become

\begin{equation}
\begin{array}{c}
\hat Q = Q \cos 2\phi -U \sin 2\phi \\
~\\ \hat U = U \cos 2\phi + Q \sin 2\phi
\end{array}
\end{equation}

\begin{figure}
\vspace{-15pt}
\centerline{\hspace{30pt}\epsfig{file=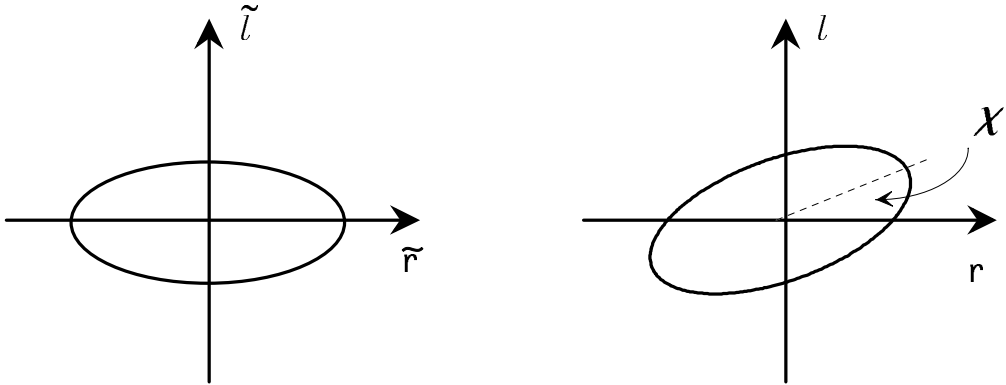,height=2in}}
\vspace{-12pt} \caption{\small The polarization ellipse}
\label{f2}
\end{figure}

\noindent When $\phi = \chi =(1/2) \arctan {\dfrac{U}{Q}} $ the axes of the
polarization ellipse coincide with the reference axes $\hat l$ and $\hat r$
(see fig.\ref{f2}).

\subsection{Electric and Magnetic Modes}

To analyze the properties of the CMB polarization it is sometimes convenient
to use rotationally invariant quantities, like the radiation intensity $I$ and
two combinations of $U$ and $Q$: $Q+iU$ and $Q-iU$. The intensity $I$ can be
decomposed into usual (scalar) spherical harmonics $Y_{lm}(\theta, \varphi)$.
\begin{equation}
I=\sum_{l,m} a_{lm} Y_{lm}(\theta, \varphi)
\end{equation}

\noindent The quantities $Q \pm iU$ can be decomposed into $\pm 2$ spin
harmonics \cite{saz96a}, \cite{saz96b}, \cite{sel97} $ Y_{lm}^{\pm 2}(\theta,
\varphi)$ \footnote{Alternatively, one can use the equivalent polynomials
derived in \cite{gol67}} :
\begin{equation}
Q \pm iU=\sum_{l,m} a_{lm}^{\pm 2} Y_{lm}^{\pm 2}(\theta, \varphi)
\end{equation}

The  $\pm 2$ spin harmonics form a complete orthonormal system (see, for
instance,  \cite{gol67},\cite{gel58}, \cite{zer70}, \cite{tho80}) and can be
written \cite{saz96b}, \cite{saz99}:
\begin{equation}
\begin{array}{c}
Y^2_{lm}(\theta, \varphi) = N^2_{lm} P^2_{lm}(\theta) e^{i m \varphi} \\
~\\ Y^{-2}_{lm}(\theta, \varphi) = N^{-2}_{lm} P^{-2}_{lm}(\theta) e^{i m
\varphi}
\end{array}
\end{equation}

\noindent where
\begin{equation}
P^s_{lm}(x) = (1 - x)^{\displaystyle (m + s) \over 2} (1 + x)^{\displaystyle
(s - m) \over 2} P^{(m + s, s - m)}_{l-s}(x)
\label{jac}
\end{equation}
\noindent is a generalized Jacobi polynomial, $s = \pm 2$ and:
$$
N^s_{lm} = \frac{1}{2^s}\sqrt{\frac{2l + 1}{4 \pi}} \sqrt{\frac{(l - s)!(l +
s)!}{(l - m)! (l + m)!}}
$$
\noindent is a normalization factor.

The harmonics amplitudes $a^{\pm 2}_{lm}$ correspond to the Fourier spectrum
of the angular decomposition of rotationally  invariant combinations of Stokes
parameters.

Because spin $\pm 2$ spherical functions form a complete orthonormal system:
\begin{equation}
\int_{4 \pi} Y^{\pm 2}_{l m}(\theta, \varphi)Y^{*\; \pm 2}_{l^\prime
m^\prime}(\theta, \varphi) d\Omega = \delta_{l m}\delta_{l^\prime m^\prime}
\label{orth}
\end{equation}

\noindent  we can write
\begin{equation}
a^{\pm 2}_{l m} =  \int_{4 \pi} d\Omega \; \; \left(Q(\theta, \varphi) \pm i
U(\theta, \varphi)\right) Y^{*\; \pm 2}_{l m} \label{ampl}
\end{equation}

Following \cite{sel97} we now introduce the so called {\it  $E$ (electric)}
and {\it $B$ (magnetic) modes} of these harmonic quantities:
\begin{equation} \begin{array}{c}
a^E_{lm} = \frac{1}{2}\left( a^{+2}_{lm} + a^{-2}_{lm}\right) \\
~\\ a^B_{lm}=\frac{i}{2}\left( a^{+2}_{lm} - a^{-2}_{lm}\right)
\end{array} \end{equation}

They have different parities. In fact when we transform the coordinate system
$Oxyz$ into a new coordinate system $\tilde O \tilde x \tilde y \tilde z$,
such that

\begin{equation}
\begin{array}{c}
\tilde  {\vec l} = \vec l \\
\tilde  {\vec r} = - \vec r
\end{array}
\end{equation}

\noindent the E and B modes transform in a similar way:
\begin{equation}
\begin{array}{c}
\tilde a^{E} = a^{E} \\
\tilde a^{B} =- a^{B}
\end{array}
\end{equation}

\noindent $Q$ remains identical in both reference systems and $U$ changes
sign.

It is important to remark that $a^E$ and $a^B$ are uncorrelated.

In terms of $Q$ and $U$ we can write:
\begin{equation} \begin{array}{c}
a^E_{l m} =\dfrac{1}{2} \int d\Omega \; \; \left(Q \left( Y^{+2}_{l m} +
Y^{-2}_{l m}\right) + iU\left( Y^{+2}_{l m} - Y^{-2}_{l m} \right) \right) \\
~\\ a^B_{l m} =\dfrac{1}{2} \int d\Omega \; \; \left(iQ \left( Y^{+2}_{l m} -
Y^{-2}_{l m}\right) - U\left( Y^{+2}_{l m} + Y^{-2}_{l m} \right) \right)
\end{array} \end{equation}

\noindent
therefore :
\begin{equation}
\left<(a^E_{l m})^2\right> - \left<(a^B_{l m})^2\right> = 2 \int d\Omega
\left(\left<|Q^2|\right> - \left<|U^2|\right>\right) \left(Y^{+2}_{l m} Y^{* -2}_{l m} + Y^{* +2}_{l m}
Y^{ -2}_{l m}\right) \label{syn1}
\end{equation}
Here $\left<|Q^2|\right>$ and $\left<|U^2|\right>$ designate values of
correlators of delta correlated 2D stochastic fields $Q$ and $U$.
Omitting mathematical details, the correlation equations  for $Q$
and $U$ are:
\begin{equation}
\begin{array}{c}
\left<Q Q^*\right> = |Q^2| \delta(\Omega - \Omega^\prime) \\
~\\ \left<UU^*\right> = |U^2| \delta(\Omega - \Omega^\prime) \\
~\\ \left<QU^*\right>= 0
\end{array}
\end{equation}
\noindent where $\delta(\Omega - \Omega^\prime) =\delta(\cos \theta -\cos
\theta^\prime) \cdot \delta(\varphi - \varphi^\prime)$ is the Dirac delta -
function on the sphere.
(In the following we will sometimes omit indexes $l$ and $m$).

\section{Synchrotron Radiation and its Polarization}

Synchrotron radiation results from the helical motion of extremely
relativistic electrons around the field lines of the galactic magnetic field
(see, for instance \cite{gin64}, \cite{gin69}, \cite{wes59}). The electron
angular velocity

\begin{equation}
\omega_e =\dfrac{e H_p}{m_ec}~\dfrac{m_e c^2}{\mathcal{E}} = \omega_o
\dfrac{m_e c^2}{\mathcal{E}} \label{gyr1}
\end{equation}

\noindent is determined by the ratio between $H_p$, the component of the
magnetic field orthogonal to the particle velocity, and $\mathcal{E}$, the
electron energy. As it moves around the magnetic field lines the electron
radiates.
\par~\par
\noindent a){\it Single electron}
\par
\begin{figure}
\vspace{-15pt}
\centerline{\hspace{30pt}\epsfig{file=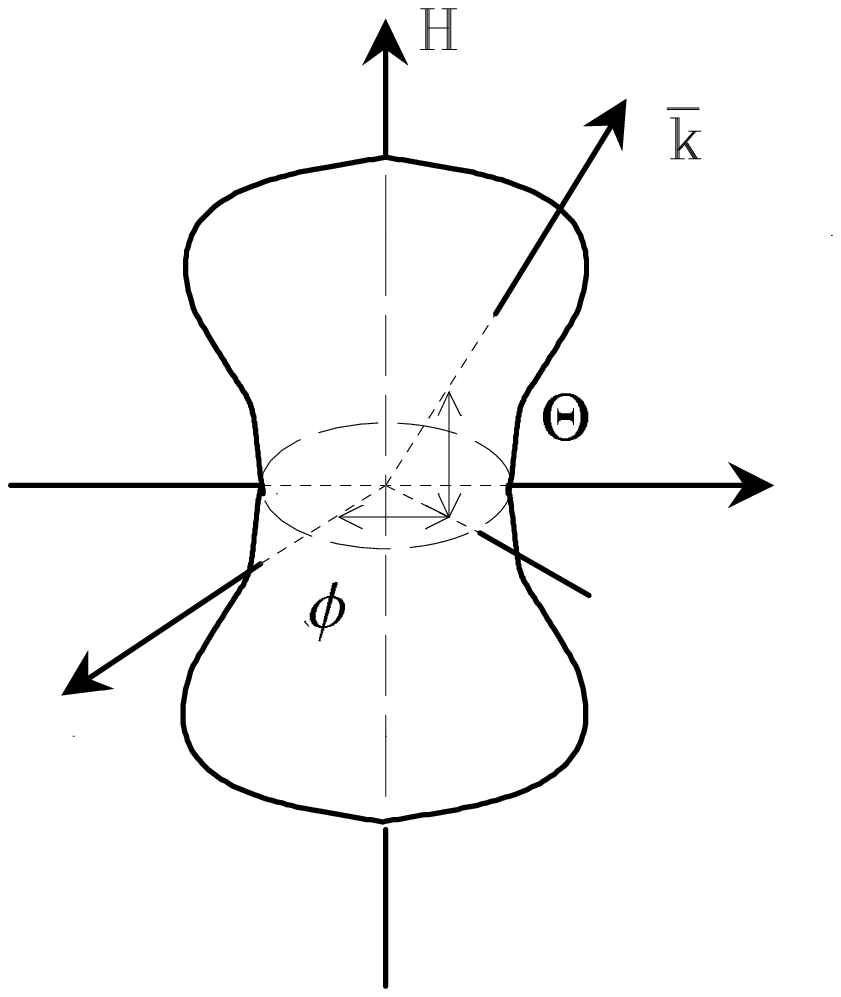,height=2in}}
\vspace{-12pt} \caption{\small Spatial distribution of the
cyclotron radiation produced by a single electron (see
eq.\ref{dumbb}) } \label{f3}
\end{figure}

\noindent Until the circular velocity is small ($v<<c$, cyclotron
radiation) the electron behaves as a rigid dipole which rotates with
gyrofrequency (\ref{gyr1}) in a plane orthogonal to the magnetic field
direction and emits a single line. The spatial distribution of the radiation
has dumbbell shape (see fig.\ref{f3}):
\begin{equation}
I(\Theta, \Phi) \sim (1 + \cos^2 \Theta)
\label{dumbb}
\end{equation}
\noindent The radiation is circularly polarized along the dumbbell axis
($\Theta = 0$) and linearly polarized in directions orthogonal to it ($\Theta
= 90^{\circ}$).

\par When the electron velocity increases the radiation field changes
until at $v \approx c$, ($\mathcal{E}$ $\gg m_e c^2$) it assumes the peculiar
charachters of synchrotron radiation:

i)radiated power proportional to $\mathcal{E}$ $^2$ and $H_p^2$,

ii)continous frequency spectrum so concentrated around:
\begin{equation}
\omega_m = \omega_o \dfrac{\sqrt{1 - \dfrac{v^2}{c^2}}}{1 -
\dfrac{v}{c}\cos\psi} ~\propto H_p\mathcal{E} ^2
\end{equation}
\par
\begin{figure}
\vspace{-15pt}
\centerline{\hspace{30pt}\epsfig{file=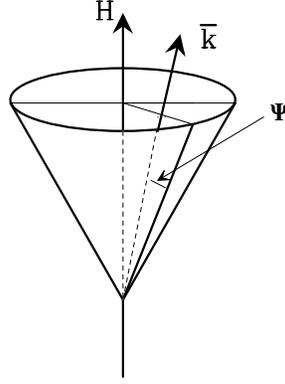,height=2in}}
\vspace{-12pt} \caption{\small Velocity cone of an
ultrarelativistic ($v\simeq c$) electron  ($H$ = magnetic vector,
$k$ wave vector, $\psi$ angle between the electron velocity and
the direction of observation)} \label{f4}
\end{figure}

\begin{figure}
\vspace{-15pt}
\centerline{\hspace{30pt}\epsfig{file=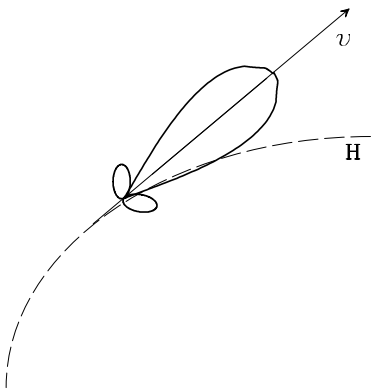,height=2in}}
\vspace{-12pt} \caption{\small Distribution of the radiation
emitted by an ultrarelativistic ($v \simeq c$) electron with
istantaneous velocity $v$ spiralling (dashed line) around the
lines of the magnetic field $H$ (orthogonal to the sheet)}
\label{f5}
\end{figure}

\noindent ($\psi$ is
the angle between the velocity vector $\vec v$ and the wave vector $\vec k$,
see fig.\ref{f4}), that in a given direction emission can be assumed monochromatic,

iii)radiation almost totally emitted in the forward direction of the
electron motion, inside a narrow cone \footnote{the symmetry
plane($\Theta = 90^{\circ}$) of the cyclotron dumbbell beam, seen by a fast
moving observer ($v \approx c$) becomes a cone folded around the direction of
movement} of aperture (see fig.\ref{f5})
\begin{equation}
\psi \approx \dfrac{m_e c^2}{\mathcal{E}}
\end{equation}
\par Inside that cone ($\cos \psi \approx 1$) the frequency is maximum and
equal to
\begin{equation}
\omega_{m,o} \approx \omega_o \left(\dfrac{{\mathcal{E}}}{m_e c^2}\right)^2,
\end{equation}
\noindent In the opposite direction ($\cos \psi \approx -1$) intensity and
frequency are sharply reduced.

iv)radiation 100 $\%$ linearly polarized when seen along the surface of the
emission cone.Inside the cone the linear polarization is still dominant but a
small fraction of circular polarization exists, ($V \sim o(\frac{m_e c^2}{E})$
and $I \sim U \sim Q$). Outside the cone the very small fraction of radiation
produced is elliptically polarized and becomes circularly polarized when seen
in directions orthogonal to the circular component of the electron spiral
motion, i.e. along the $H_p$ direction. So the Stokes parameters depend on
$H$, the angle $\mu$ between the $H$ and the line of sight, and the
dimensionless frequency $\nu/\nu_c$, where $\nu_c = 1.5(\omega_{m,o}/2\pi)$ is
the so called {\it critical frequency}, \cite{gin64},\cite{gin69}, itself
function of $H$ (see eq.\ref{gyr1}).

\par~\par
\noindent b){\it Cloud of monoenergetic electrons}
\par\noindent When the effects of many monoenergetic electrons with
uniform distribution of pitch angles are combined, $I, Q$ and $U$ are
reinforced (the Stokes parameters are additive) while $V$ is erased. In fact

\begin{equation}
\begin{array}{c}
I(\nu) =c_1 H_p F(\dfrac{\nu}{\nu_c}) \\
~\\ Q(\nu)=c_2H_pF_p(\dfrac{\nu}{\nu_c}) \cos 2\chi  \\
~\\ U(\nu)=c_2H_pF_p(\dfrac{\nu}{\nu_c}) \sin 2\chi  \\
~\\ V(\nu) \approx 0
\end{array}
\end{equation}

\noindent where $c_1$ and $c_2$ are constants, $F$ and $F_p$ frequency
functions (almost monochromatic), and $\chi$ the angle between the projection
of the magnetic vector on the observer plane and an axis on that plane (The
projection of the magnetic vector on the observer plane is the
minor axis of the polarization ellipse). When the magnetic field direction
varies, the angle $\chi$ varies therefore $\cos 2\chi$ and $\sin 2\chi$ must
be averaged on the magnetic field distribution along the line of sight through
the cloud of emitting electrons. In conclusion the degree of polarization $p$
varies between a maximum value, (uniform magnetic field) and zero (magnetic
field randomly distributed).
\par~\par
\noindent c){\it Electrons with power law energy spectrum}

In the interstellar medium the radiating electrons are the cosmic ray
electrons whose spectrum is a power law energy spectrum (see for instance
\cite{ele1}, \cite{ele2}) and references therein):
\begin{equation}
N(\mathcal{E}) = K \mathcal{E}^{- \gamma}
\end{equation}

\noindent where $\gamma \approx 2.4 - 3.0$ is the spectral index,  and $K$ a
normalization constant. Because the radiation produced by each electron is
practically monochromatic the resulting radiation spectrum is also a power
law:
\begin{equation}
I(\nu) = I_0(\gamma) H_p^{\frac{\gamma +1}{2}} \nu^{-\beta'}
\label{spec1}
\end{equation}
\noindent where $\beta' = \frac{\gamma-1}{2}$ is the intensity spectral index,
$\beta = \beta' + 2$ the temperature spectral index and $I_0$ is a slow
function of $\gamma$ \cite{gin64}. When the magnetic field is not uniform
$H_p$ is replaced by $<H_p>$ and $I_0$ by a slightly different function of
$\gamma$.

\noindent The Stokes parameters, products of the intensity $I(\nu)$, the
degree of polarization $p$ and $\cos 2\chi$ or $\sin 2\chi$, are:
\begin{equation}
\begin{array}{c}
Q= c_3~H_p^{\frac{\gamma + 1}{2}} ~\nu^{-\frac{\gamma -1}{2}} \cos 2 \chi \\
~\\U=c_3~H_p^{\frac{\gamma + 1}{2}} ~\nu^{-\frac{\gamma-1}{2}} \sin 2 \chi \\
~\\V \approx 0 \label{spec2}
\end{array}
\end{equation}

\noindent Here $c_3$ is a constant determined by $\gamma$ and by the magnetic
field distribution along the integration path. Faraday rotation, induced by
the combinations of thermal electrons (if present) and magnetic field along
the line of sight through the synchrotron emitting region, usually reduces the
polarization level. Moreover when the field is not uniform we have to use
$<\sin 2\chi>$ and $<\cos 2\chi>$ instead of $\sin 2\chi$ and $\cos 2\chi$ so
the degree $p$ of polarization decreases. Summarizing $p$ varies between 0
(random magnetic field distribution) and:
\begin{equation}
p_{max} = \dfrac{3\gamma + 3}{3\gamma + 7} < 1
\label{degreep}
\end{equation}
\noindent (uniform magnetic field and Faraday rotation absent).

When one looks in different directions through the interstellar medium $H$,
$H_p = H\sin \mu$, $\chi$, $\gamma$, $\beta$ and $K$ vary (see for instance
\cite{wie01}, \cite{rei01}. Practically their values, averaged along the line
of sight, behave as random variables \footnote{e.g. $F= \left( H\sin
\mu\right)^{\frac{\gamma +1}{2}}$ is a nonzero mean, nonzero variance
variable; $\cos 2 \chi$ and $\sin 2 \chi$ are zero mean, nonzero variance
variables}  therefore the Stokes parameters $Q_s$ and $U_s$ ($s$ stays for
synchrotron) associated to the galactic (synchrotron) foreground behave as
stochastic functions of the direction of observation. We can therefore write:

\begin{equation}
\left<Q^2_s\right> = \left<U^2_s\right>
\label{main}
\end{equation}

It follows that for synchrotron radiation if $Q_s \ne 0$ also $U_s \ne 0$,
therefore both electric and magnetic modes exist:
\begin{equation} a^{E,s} \ne
0, \hskip2cm a^{B,s} \ne 0
\end{equation}

\noindent moreover for the synchrotron
background we can write (see eqs. (\ref{syn1}) and  (\ref{main})):
\begin{equation}
\left<(a^{E,s})^2\right> = \left<(a^{B,s})^2\right> \label{syn}
\end{equation}

\section{CMB and its Stokes Parameters}

In a homogeneous and isotropic Universe the only quantities which
change as the Universe expand are temperature and intensity $I=I_l
+ I_r$: both decrease adiabatically. Because
this is true for $I_l$ and $I_r$ separately, we do not expect anisotropy nor
polarization therefore $Q=0$ and $U = I_u =0$ is a natural consequence.

On the contrary, inhomogeneities and perturbations of the matter density or
of the gravitational field, induce anisotropy and polarization of the CMB. At
the recombination epoch linear polarization is produced by the Thomson
scattering of the CMB on the free electrons of the primordial plasma. The
polarizarion tensor can be calculated solving the Boltzman equations, which
describe the transfer of radiation in a nonstationary plasma permeated by a
variable and inhomogeneous gravitational field \cite{bas80}, \cite{saz84},
\cite{har93}, \cite{saz95}.

In our Universe the gravitational field can be divided in two parts: a
background field, with homogeneous and isotropic FRW metric, and an
inhomogeneous and variable mix of waves: density
fluctuations, velocity fluctuations, and gravitational waves. Because of their
transformation laws these waves are also said scalar, vector and tensor
perturbations, respectively.

Scalar (density) perturbations affect the gravitational field, the density of
matter and its velocity distribution. They were discovered by astronomers who
studied the matter distribution in our Universe on scale from $\sim 1$ Mpc to
$\sim 100$ Mpc. It is firmly believed they are the seeds of the large scale
structure of the Universe and are reflected by the large scale CMB anisotropy
detected for the first time at the beginning of the `90s \cite{str92},
\cite{smo92}. Their existence is predicted by the great majority of models of
the early Universe.

Vector perturbations, associated to rotational
effects, perturb only velocity and gravitational field. They are not predicted
by the inflation theory and it is common believe that they do not contribute to
the anisotropy and polarization of the CMB.

Tensor perturbations affect exclusively the gravitational field.

In the following we will concentrate our attention on density waves, the
dominant perturbation. Because observation shows that the effects of these
perturbations are small, we can treat them as small variations $\delta_l$,
$\delta_r$, $\delta_u$ of $I_l$, $I_r$ and $I_u$.  If we introduce the
auxiliary functions $\alpha$ and $\beta$:
\begin{equation}
\begin{array}{c}
\delta_l + \delta_r = (\mu^2 - {1 \over 3}) \alpha, \\
~\\ \delta_l - \delta_r = ( 1 - \mu^2) \beta, \\
~\\ \delta_u =0,
\end{array}
\end{equation}

\noindent ($\mu$ is the angle between the line of sight and the wave vector)
and assume plane waves, the Boltzman equations (see for instance
\cite{saz95}, \cite{saz96a},  \cite{saz96b} and reference therein) become:

\begin{equation}
\begin{array}{c}
\dfrac{d \alpha}{d \eta} = F - \dfrac{9}{10} \sigma_T n_e a(\eta) \alpha -
\dfrac{6}{10} \sigma_T n_e a(\eta) \beta \\
~\\ \dfrac{d \beta}{d \eta} = - \dfrac{1}{10} \sigma_T n_e a(\eta) \alpha -
\dfrac{4}{10} \sigma_T n_e a(\eta) \beta
\end{array}
\end{equation}

\noindent where $F$ is the gravitational force which drives both anisotropy
and polarization, $\sigma_T$ is the Thomson cross-section, $n_e$ is the
density of free electrons, and $a(\eta)$ the scale factor.

These equations give:
\begin{equation}
\begin{array}{c}
 Q= -\frac{1}{7}(1 - \mu^2) \int F(\eta) \left( e^{-\tau} - e^{-{3 \over
10} \tau}\right) d\eta \\
~\\ U= 0
\end{array}
\end{equation}

\noindent where $\tau(\eta)$ is the optical depth of the region where the
phenomenon occurs. Rotating the coordinate system we can generate a new pair
of Stokes parameters ($Q^\prime ,U^\prime$): no matter which is the system of
reference we choose these parameters satisfy the symmetry parity condition.

Because as we saw for the CMB there is a system in which $Q \ne 0$
and $U=0$, we may conclude that magnetic modes of the CMB polarization vanish
and only electric modes exist \cite{sel97}. Therefore for the CMB we can
write:
\begin{equation}
a^{E,d} \ne 0 \hskip2cm a^{B,d}=0
\label{aed}
\end{equation}
where index $d$ stays for density perturbation.

Concluding: in our approximation the CMB  polarization is a random tensor
field whose characteristcs are set by primordial density perturbations (see,
for instance \cite{dol90}).

\section{Separation of the polarized components of Synchrotron and CMB Radiation}
The CMB radiation we receive is a diffuse background which reaches us mixed
with foregrounds of local origin. When the anisotropy of the CMB was
detected, to remove the foregrounds from maps of the diffuse radiation, data
were reorganized in the following way:
\begin{equation}
\hat T_d(x,y) =  \hat T_n(x,y) + \hat M_i(x,y) \hat T_{i,c}(x,y)
\end{equation}

\noindent where $\hat T_d$, is a two dimension vector  (map) which gives the
total signal measured at different points $(x,y)$ on the sky, $\hat T_n$ the
noise vector, $\hat M$ the matrix which combines the components
$\hat T_{i,c}$ of the signal. At each point $(x,y)$ we can in fact write:
\begin{equation}
T_d = g_nT_n + g_{cmb}T_{cmb, c} + g_{syn}T_{syn, c} + g_{ff}T_{ff, c}
+g_{dust}T_{dust, c} +...
\end{equation}

\noindent where $g_i$ are weights, given by $M$, $T_{syn, c}$ is the
synchrotron component, $T_{ff, c}$ the free-free emission component,
$T_{dust, c}$ the dust contribution and so on. Using just one map the signal
components cannot be disentangled. If however one has maps of the same region
of sky made at different frequencies it is possible to write a system of
equations. Provided the number of maps and equations is sufficient, the
system can be solved and the components of $T_d$ separated, breaking the
degeneracy. We end up with a map of $T_{cmb}$ which can be used to estimate
the CMB anisotropy.

When we look for polarization at each point on the sky we measure tensors
components instead of scalar quantities, therefore to disentangle
the polarized components of the CMB we need a greater number of equations.
Here we will concentrate on the separation of the two dominant components
of the polarized diffuse radiation: galactic synchrotron (plus dust) foreground
and CMBP.

\subsection{The estimator $D$}
Instead of observing the same region of sky at many
frequencies, we suggest a different approach. It takes advantage of the
differences between the statistical properties of the two most important
components of the polarized diffuse radiation (CMB (background) and
synchrotron (foreground) radiation) and does not require multifrequency maps.
\par~\par

\noindent We define the estimator:
\begin{equation}
D = \left<(a^E)^2\right> - \left<(a^B)^2\right>
\end{equation}
\noindent where $a^E = a^{E,s} + a^{E,d}$ and $a^B = a^{B,s} + a^{B,d}$ (here
and in the following indexes $s$ or $d$ stay for synchrotron and density
perturbations, respectively).
\par \noindent Because
\begin{itemize}

\item $E$ and $B$ modes of synchrotron do not correlate each other neither
correlate with the CMB modes

\noindent  $\left<(a^E)^2\right> = \left<(a^{E,s})^2\right> +
\left<(a^{E,d})^2\right> + 2\left<a^{E,s} a^{E,d}\right> =
\left<(a^{E,s})^2\right> + \left<(a^{E,d})^2\right>$,

\noindent $\left<(a^B)^2\right> = \left<(a^{B,s})^2\right> +
\left<(a^{B,d})^2\right> + 2\left<a^{B,s} a^{B,d}\right> =
\left<(a^{B,s})^2\right> + \left<(a^{B,d})^2\right>$,

\item Electric and Magnetic modes of the CMB polarized component satisfy the
condition
$$
a^{E,d} \ne 0 \hskip4cm a^{B,d} =0
$$

\item Electric and Magnetic modes $a^{E,s}$ and  $a^{B,s}$, of the synchrotron
radiation are both different from zero and their average values identical,
\end{itemize}

\par\noindent we get
\par~\par
\begin{equation}
D = \left<(a^{E,d})^2\right>
\end{equation}

\par~\par\noindent It means that $D$ provides an estimate of the
contribution of the CMB to maps of the polarized diffuse radiation
contaminated by galactic synchrotron emission.

Let` s now consider the angular power spectrum of $D$. For multipole $l$ we
can write:
\begin{equation}
D_l = (a^E_l)^2 - (a^B_l)^2 = \dfrac{1}{2l + 1} \sum_{m = -l}^l \left(|a^E_{l
m}|^2 - |a^B_{l m}|^2 \right) \label{dest}
\end{equation}

\noindent where $|a^E_{l m}|^2$ and $|a^B_{l m}|^2$ are random variables with
gaussian distribution $p(a_{lm}^{E,B})$ (see eqs.(\ref{gauss1}) and
(\ref{gauss2}) in Appendix A). According the ergodic theorem (in the limit of
infinite maps, the average over 2D space is equivalent to the average over
realisations) the average value of $D_l$ is equal to the difference of the
average values of $|a^E_{l m}|^2$ and $|a^B_{l m}|^2$ summed over $m$. Taking
into account equation (\ref{gauss2}) we can therefore write:
\begin{equation}
\left< D_l \right> =  \left(|a^E_{l}|^2 -  |a^B_{l}|^2 \right) \label{D1}
\end{equation}

\noindent where \footnote {Equation (\ref{gauss0}) is an explicit form of the
average of the stochastic variables $|a_{lm}^E|^2$ and $|a_{lm}^B|^2$ over a
probability density $p(a_{lm}^{E,B})$, the short form being triangle
brackets}:

\begin{equation} \begin{array}{c}
(a^E_l)^2 = \int\limits^{\infty}_{-\infty}|a_{lm}^E|^2 p(a_{lm}^E) d^2
a^E_{lm} = \left< |a_{lm}^E|^2\right>
  \\
~\\
(a^B_l)^2 = \int\limits^{\infty}_{-\infty}|a_{lm}^B|^2 p(a_{lm}^B) d^2
a^B_{lm} = \left< |a_{lm}^B|^2\right>
\label{coco}
\end{array}
\label{gauss0}
\end{equation}

\noindent Comparing eq.\ref{coco} with the ordinary definition of multipole
coefficients:
\begin{equation}
C_l^{E,B}= \dfrac{1}{2l + 1} \sum_{ m= -l}^{ l} \left< |a_{lm}^{E,B}|^2\right>
\label{multipole}
\end{equation}
\noindent  it immediately appear that we can write:
\begin{equation} \begin{array}{c}
 (a^E_l)^2 = C^E_l \\
~\\
(a^B_l)^2 = C^B_l
 \label{equal}
\end{array}
\end{equation}

\subsection{Separation uncertainty}
For synchrotron radiation $\left< D_l \right>$ should be zero, non zero for
density perturbations, but on real maps it is always different from zero. In
fact a map is just a realization of a stochastic process and the amplitudes
of  $|a^E_{l m}|^2$ and $|a^B_{l m}|^2$, averaged over 2D sphere, have
uncertainties which add quadratically making $\left< D_l \right> \ne 0$ even
in the case of synchrotron polarization. This effect, very similar to the
well known "cosmic variance" of anisotropy   \cite{var}, \cite{saz95b} (the
real Universe is just a realisation of a stochastic process, therefore there
will be always a difference between the realization we measure and the
expectation value) does not vanish if observations are repeated.

\par~\par The variance of $D_l$ is
\begin{equation}
{\bf V} (D_l^2) = \left< D_l^2 \right> - \left< D_l \right>^2
\end{equation}

Being sums over $m, ~(-l\le m \le +l)$ of $2l+1$ stochastic values with
gaussian distribution, $a^E$ and $a^B$ have $\chi^2$ distributions with
$2*(2l +1)$ degrees of freedom, therefore we expect their variances can be
written as

\begin{equation}
\delta (a^{E, B})^2 \propto \sqrt{\dfrac{2}{2 l + 1}}
\label{var1}
\end{equation}

\noindent More explicitly
\begin{equation}
\left< D^2_l \right> = \dfrac{1}{2 l +1}  \left((a_l^E)^4 +  (a_l^B)^4 \right)
\label {Dl20}
\end{equation}
\noindent and when the synchrotron foreground is dominant ($a_l^E=a_l^B$)
\begin{equation}
\left< D^2_l \right> = \dfrac{2}{2 l +1} (a^E_l)^4 \label{Dl21}
\end{equation}

\noindent in agreement with (\ref{var1}).

\subsection{A criterium for CMBP detection}
The synchrotron foreground is a sort of {\it system noise} which hampers
detection of the {\it signal}, the CMB polarization. At frequencies
sufficiently high ($\approx$ above 50 GHz, as we will see in the next
section) the noise is small compared to the signal therefore direct detection
of CMBP is possible. At low frequencies on the contrary the CMBP signal is
buried in the synchrotron plus dust noise. In this case to recognize the
presence of the CMB polarization we can use our estimator $D$.

At angular scale $l$, to be detectable the CMBP must satisfy the condition

\begin{equation}
C_l^{E,d} > A \cdot C_l^{E,s} \label{sn1}
\end{equation}

\noindent where $C_l^{E,i}$ are the coefficients of the multipole expansion of
the $E$ modes and $A$ is the confindence level of the signal detectability.
In a similar way we can write for our estimator:
\begin{equation}
D_l^{E,d} > A \cdot D_l^{E,s} \label{sn2}
\end{equation}
\noindent where, as we said before, $D_l^{E,d} = C_l^{E,d}$ and
\begin{equation}
D_l^{E,s} = \sqrt{\dfrac{2}{2l+1}} (a_l^{E,s})^2 \label{sn3}
\end{equation}

Therefore, the criterium for the CMBP detection becomes
\begin{equation}
D_l^{E,d} \geq A \cdot \sqrt{\dfrac{2}{2l+1}} C_l^{E,s} \label{sn4}
\end{equation}

\noindent This means that using D we can recognize CMBP in a map of the
polarized diffuse radiation with an uncertainty ~$\simeq
\sqrt{\dfrac{2}{2l+1}} C_l^{E,s}$ which decreases as $l$ and angular
resolution increases.

\section{Angular power spectra of polarized synchrotron.}

The angular power spectra of the polarized component of the synchrotron
radiation have been discussed by \cite{bur01}, \cite{bur02}, \cite{tuc00},
\cite{tuc02} and \cite{brus02}. It appears that the power spectra of the
degree of polarization $p$ and of electric and magnetic modes $E$ and $B$
follow power laws up to $l \sim 10^3$. The spectral index of the power law
which holds for the degree of polarization is $\alpha_p \simeq 1.6 - 1.8$.
For the $E$ and $B$ modes different authors get from the observations values
of the spectral index different, but marginally consistent. In paper
\cite{tuc02} and \cite{brus02}, using Parkes data, the authors get

\begin{equation}
C^{E,B}_l \sim C_{0, E, B} \left(\dfrac{l_0}{l}\right)^{\alpha_{E,B}}
\label{syn111}
\end{equation}

\noindent  with $\alpha_E \approx \alpha_B \approx 1.4 \div 1.5$ and
dependence of $\alpha$ on the sky region and the frequency.

In paper \cite{bur01}, using Effelsberg and Parkes data, the authors get:
\begin{equation}
C^{E,B}_l = C_0 \cdot 10^{-10}  \left(\dfrac{450}{l}\right)^{\alpha}
\cdot  \left(\dfrac{2.4 \mbox{GHz}}{\nu}\right)^{2\beta}
\label{synEBB}
\end{equation}
with  $\alpha =(1.8 \pm 0.3)$, $\beta = 2.9$ and $C_0= (1.6 \pm 1)$ (here we
modified the original expression given in \cite{bur01} writing it in
adimensional form). As said above the spectral indexes $\alpha$ obtained by
\cite{tuc02} \cite{brus02} and \cite{bur01} are marginally consistent.

Both authors got their results analyzing low frequency data (1.4, 2.4 and 2.7
GHz), therefore the extension of their spectra to tens of GHz, the region
where CMB observations are usually made, depend on the accuracy of $\beta$,
the temperature spectral index of the galactic synchrotron radiation. A
common choice is $\beta=2.9$ but in literature there are values of $\beta$
ranging between $\approx 2.5$ and $\approx 3.5$. Moreover $\beta$ depends on
the frequency and the region of sky where measurements are made (see
\cite{ele1}, \cite{index}, \cite{zanno}, \cite{bersa}). Finally the values of
$\beta$ in literature have been obtained measuring  the total (polarized plus
unpolarized) galactic emission. In absence of Faraday effect $\beta ,
\beta_{pol}$ and $\beta_{unpol}$, the spectral indexes  of the total,
polarized and unpolarized components of the galactic emission, are identical
(see eqs. (\ref{spec1}),(\ref{spec2})). However when Faraday effect with its
$\nu^{-2}$ frequency dependence is present, we expect that, as frequency
increases, the measured value of the degree of polarization (se eq.
(\ref{degreep})) increases, up to
$$
p \leq p_{max} = \dfrac{3\beta -3}{3\beta -1}
$$

\noindent therefore we should measure $\beta_{pol} \leq \beta$. The expected
differences are however well inside the error bars of the available data
therefore at present we can neglect differences and set $\beta \simeq
\beta_{pol} \simeq \beta_{unpol}$.

Instead of extrapolating low frequency results it would be better to look for
direct observations of the galactic emission and its polarized component at
higher frequencies. Unfortunately above 5 GHz observations of the galactic
synchrotron spectrum and its distribution are rare and incomplete.
  At 33 GHz observations by \cite{dav99} give for some patches on the sky
a galactic temperature of about $2 \div 4 \mu K$ from which follows that at
the same frequency we can expect polarized foreground signals up to several
$\mu K$. At 14.5 GHz observations made at OVRO \cite{muk02} give synchrotron
signals of 175 $\mu K$, equivalent to about 15 $\mu K$ at 33 GHz of which up
to 10 $\mu K$ can be polarized.

In conclusion there are large uncertainties when one try to decide at which
frequency the polarized component of the galactic diffuse emission does not
disturb observation of the CMB. To be on the safe side we can say, and all the
authors of papers \cite{bur01}, \cite{bur02}, \cite{tuc00}, \cite{tuc02} and
\cite{brus02} agree, that the CMB polarized component surely overcomes the
polarized synchrotron foreground only above 50 GHz, therefore direct
observations of the CMB are better made at frequencies greater of 50 GHz.
Below 50 GHz observations of CMBP are problematic because the contamination
by galactic polarized emission can be high and its evaluation, usually by
multifrequency observations, not sure.

\begin{figure}
\vspace{-15pt}
\centerline{\hspace{30pt}\epsfig{file=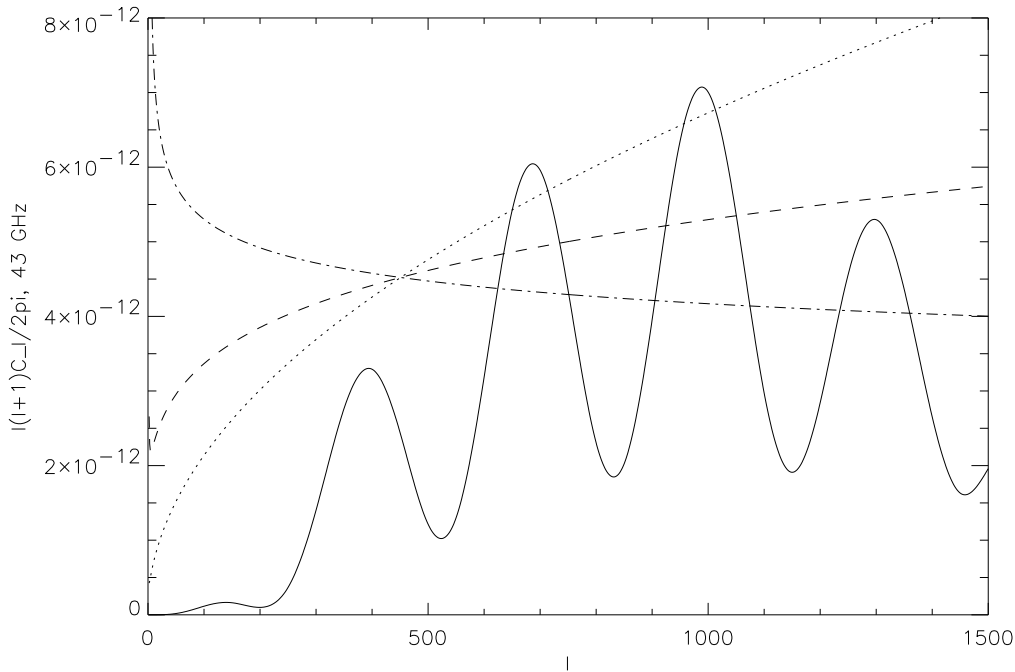,height=3.5in}}
\vspace{-12pt} \caption{\small Power spectrum versus multipole
order $l$ of the polarized components of CMB (solid line) and
three possible power spectra of the galactic synchrotron radiation
($\alpha = 1.5$ (dotted line), $\alpha = 1.8$ (dashed line) and
$\alpha = 2.1$ (dot-dash line)) calculated at 43 GHz. (See text
for details of calculations and model)} \label{f9}
\end{figure}

This conclusion is supported by figure \ref{f9}. It shows plots versus the
multipole order $l$ of the power spectra of the polarized components of CMB
and galactic synchrotron calculated at 43 GHz. The CMB spectrum has been
obtained using CMBFAST \cite{zal02} and standard cosmological conditions
(CMB power spectrum normalized to the COBE data at low $l$, $\Omega_b =
0.05$, $\Omega_{CDM}=0.3$ , $\Omega_{\Lambda} = 0.65$, $\Omega_{\nu} =0$,
$H_0 =65$ km/sec/Mpc, $T_{CMB}=2.726 K$ , $Y_{He} =0.24$, standard
recombination). The synchrotron spectrum has been calculated assuming for
$E_l$ and $B_l$ the scaling law (\ref{synEBB}) with $C_0=2.6$ (most
pessimistic case), and $\alpha = 1.5;~ 1.8;~ 2.1$ and $\beta =2.9$
respectively. At this frequency (43 GHz) the CMB power becomes comparable to
the synchrotron power only at very small angular scales ($l \geq 500$).

 Similar calculations at other frequencies
confirm that only above $\simeq 50$ GHz and at small angular scales (large
values of $l$) the CMBP power spectrum overcomes the synchrotron spectrum.
Below $\simeq 50$ GHz the CMBP power is always below the power level of
synchrotron  therefore CMBP detection is impossible even at small angular
scale.
\par~\par
We can however overcome this limit and plan CMBP observations at lower
frequencies if we use our estimator. To see how it improves the CMBP
detectability we calculate the angular power spectrum of $D^s_l$, the
estimator we expect when the diffuse radiation is completely dominated by
synchrotron galactic emission. Combining eqs.(\ref{multipole}), (\ref{Dl21})
and (\ref{synEBB}) we can write:

\begin{equation}
\sqrt{\left< (D^s_l)^2 \right>} =C_0 \cdot 10^{-10} \sqrt{\dfrac{2}{2l +1}}
\left(\dfrac{450}{l}\right)^{\alpha} \cdot  \left(\dfrac{2.4
\mbox{GHz}}{\nu}\right)^{2 \beta} \label{synD}
\end{equation}

This quantity can be directly compared with the power spectrum of CMBP
because as we saw (see eqs.(\ref{aed}), (\ref{D1}), (\ref{equal}) and
(\ref{Dl20}), $D^d_l$ the power spectrum of the estimator evaluated when the
sky is dominated by CMBP coincides with the CMBP power spectrum.

\begin{figure}
\epsfig{file=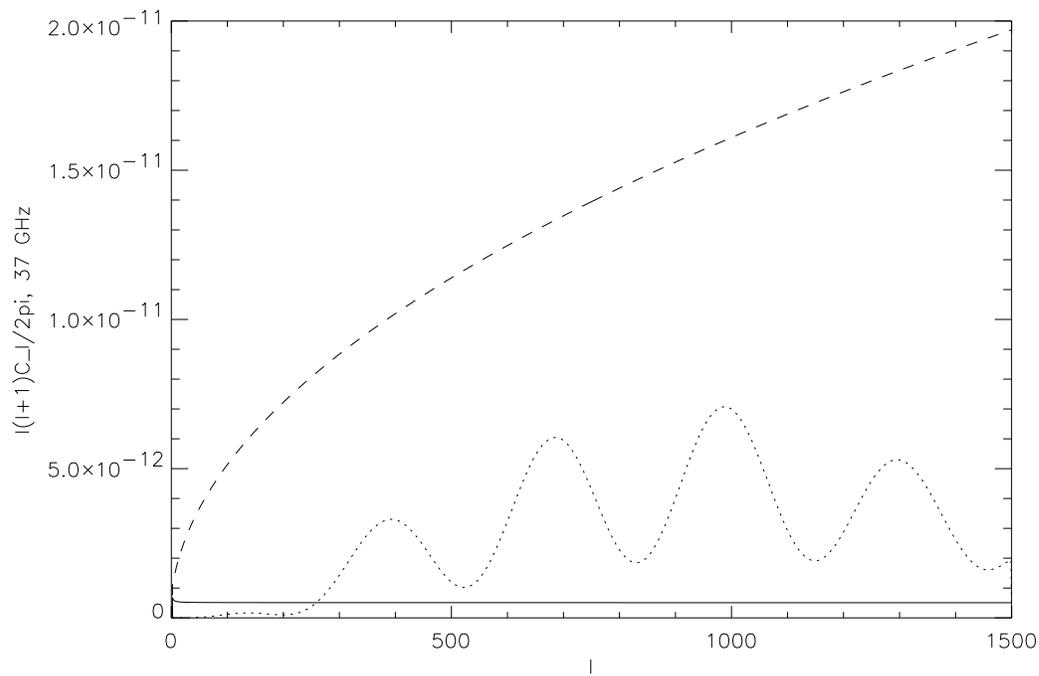,height=3.5in} \caption{\small Power spectrum
at 37 GHz of the expected value of the estimator  for a galactic
synchrotron dominated sky (solid line). The dotted line represents
both the estimator and the CMBP power spectrum for a CMBP
dominated sky. The dashed line is the power spectrum  of the
polarized component of the galactic synchrotron radiation (dashed
line) (see text for details of model and calculations)}
 \label{fD}
\end{figure}

Figure \ref{fD} shows the power spectra at
37 GHz of i)$D^s_l$ the estimator for a galactic synchrotron dominated sky
(solid line), calculated using eq.(\ref{synEBB}), ii)$D^d_l$ the estimator
for a CMBP dominated sky which coincides with the power spectrum of CMBP
(dotted line) calculated as in figure \ref{f9} with CMBFAST using the same
standard cosmological conditions, iii)the power spectrum of the polarized
component of the galactic synchrotron radiation (dashed line), calculated
using eq.(\ref{synEBB}). As expected at 37 GHz the CMBP power level is well
below the level of synchrotron power, therefore direct observations of CMBP
are impossible (the maximum value of the CMBP power is about 2.5 times smaller
than the synchrotron power at the same $l$). However above $l\simeq 250$ the
power associated to the synchrotron estimator is definitely below the power
associated to the CMBP estimator  with a maximum ratio CMBP/D $\sim 7$ at $l
= 1000$. This confirms that the use of $D$ allows to recognize the CMBP also
at frequencies well below 50 GHz.

\section{Simulation}

To further test the capability of our estimator we studied the separation of
CMB and galactic synchrotron during observations using, instead of the
expected values of $D_l$, as we did in figure \ref{fD}, measured values
simulated by random series of numbers, with gaussian distribution $\sim
N(0,1)$, zero mean and unity variance (see eqs (\ref{gauss1}) -
(\ref{gauss2})). To simulate measurements we generate $2l+1$ random numbers
to represent $a^E_{lm}$ and $2l + 1$ random numbers to represent $a^B_{lm}$.
We then use these values to work out $D_l$ using eq.(\ref{dest}).

\begin{figure}
\vspace{-15pt}
\centerline{\hspace{30pt}\epsfig{file=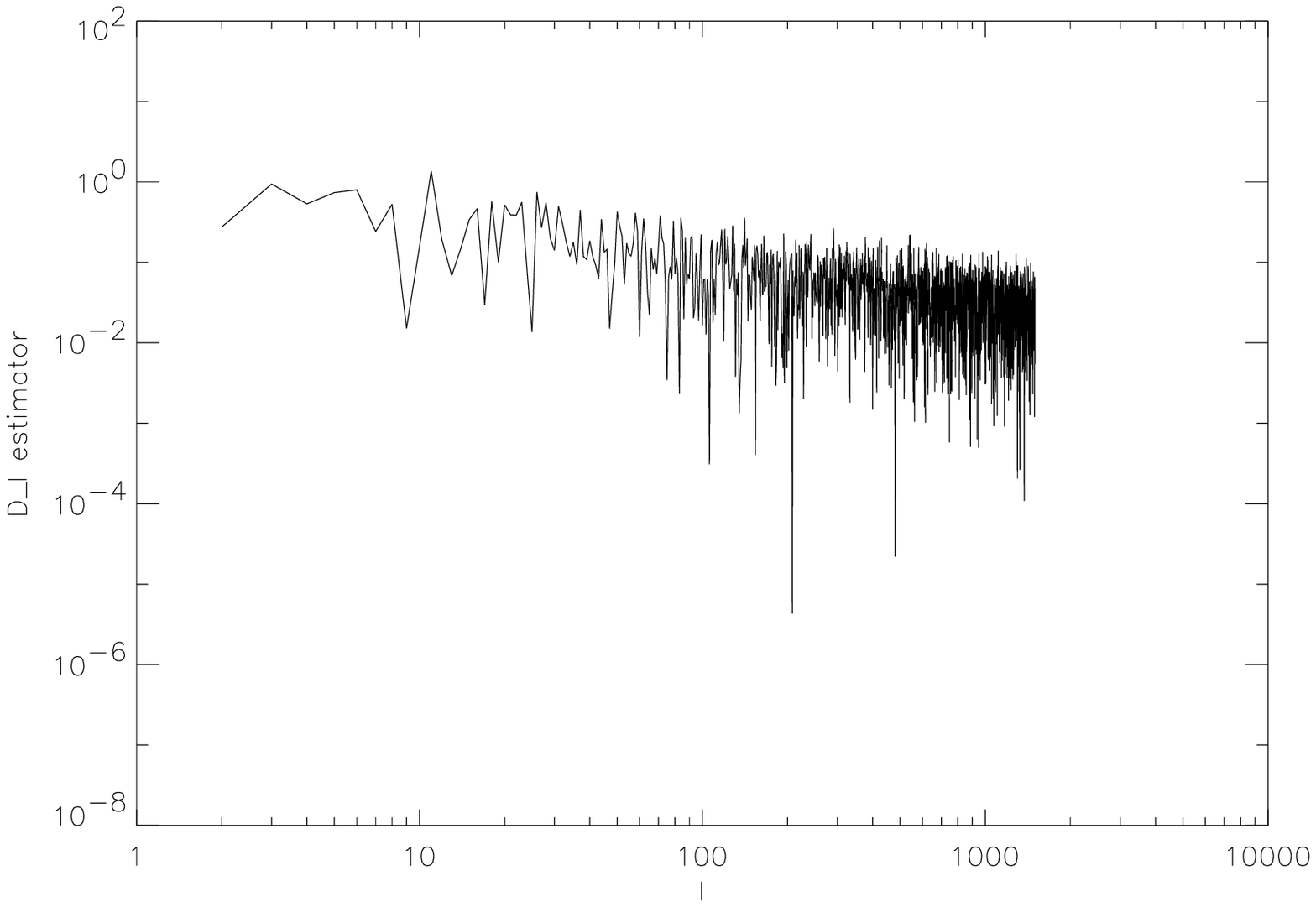,height=3.5in}}
\vspace{-12pt} \caption{\small Multipole power spectrum of
simulated measurements of the estimator $D$ for a synchrotron
dominated sky (see text) with infinite angular resolution (no
smoothing on $l$, ($\Delta l = 1$))} \label{f6}
\end{figure}
Because of their finite angular resolution observations include an average of
the signal on regions whose angular extension cover a finite multipole
interval $\Delta l$. For instance Boomerang data come from regions whose
angular extension is equivalent to $\Delta l \sim 100$ \cite{boo00},
\cite{boo01a}, \cite{boo01b}.  To get more realistic data we averaged $D_l$
on an interval $\Delta l = n$:
\begin{equation}
 \hat D_l = \dfrac{1}{n+1} |\sum\limits_{k = 0}^{k = +n} D_{l-n/2+k}|
\end{equation}

\begin{figure}
\vspace{-15pt}
\centerline{\hspace{30pt}\epsfig{file=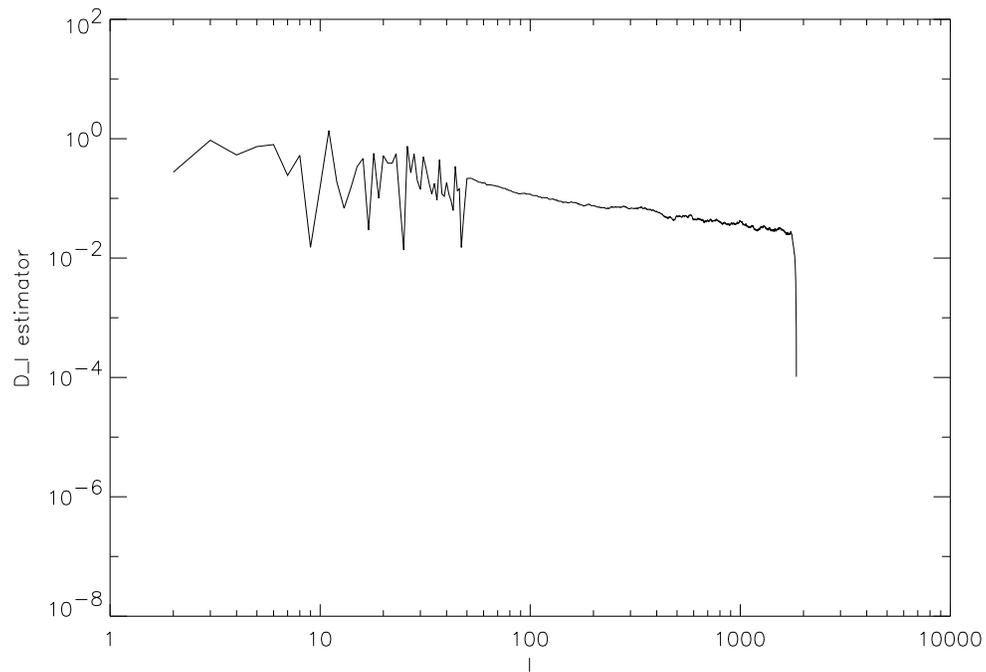,height=3.5in}}
\vspace{-12pt} \caption{\small Same as figure \ref{f6} with finite
angular resolution (smoothing on $\Delta l = 100$)} \label{f8}
\end{figure}

Figure \ref{f6},  figure \ref{f7} and figure \ref{f8} give plots of $ |\hat
D_l|$ (the sign of $D_l$ is arbitrary) versus $l$, for $\Delta l =1$, $\Delta
l =10$ and $\Delta l = 100$ respectively: the very large fluctuations of the
estimator one observes when $\Delta l =1$, above $l \simeq 400$ are
drastically reduced as soon as $\Delta l$ increases. The smoothing effect of
the average can be appreciated comparing figure \ref{f6}, figure \ref{f7} and
figure \ref{f8} where $\hat D_l$ is plotted for $n=1$, $n= 10$ and $n= 100$
respectively:

\begin{figure}
\vspace{-15pt}
\centerline{\hspace{30pt}\epsfig{file=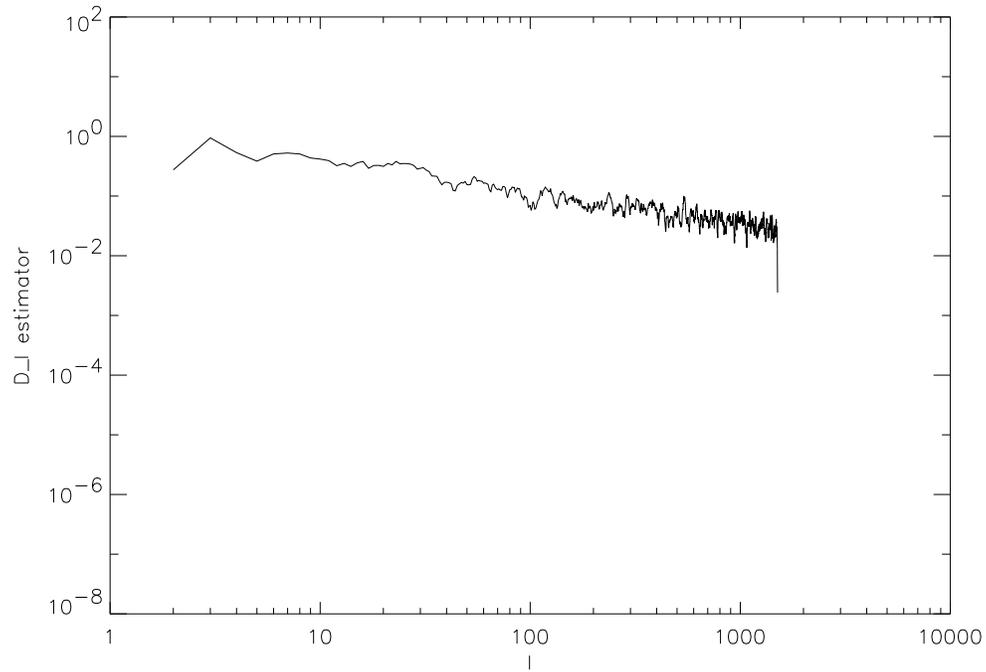,height=3.5in}}
\vspace{-12pt} \caption{\small Same as figure \ref{f6} with finite
angular angular (smoothing on $\Delta l = 10$)} \label{f7}
\end{figure}

\begin{figure}
\vspace{-15pt}
\centerline{\hspace{30pt}\epsfig{file=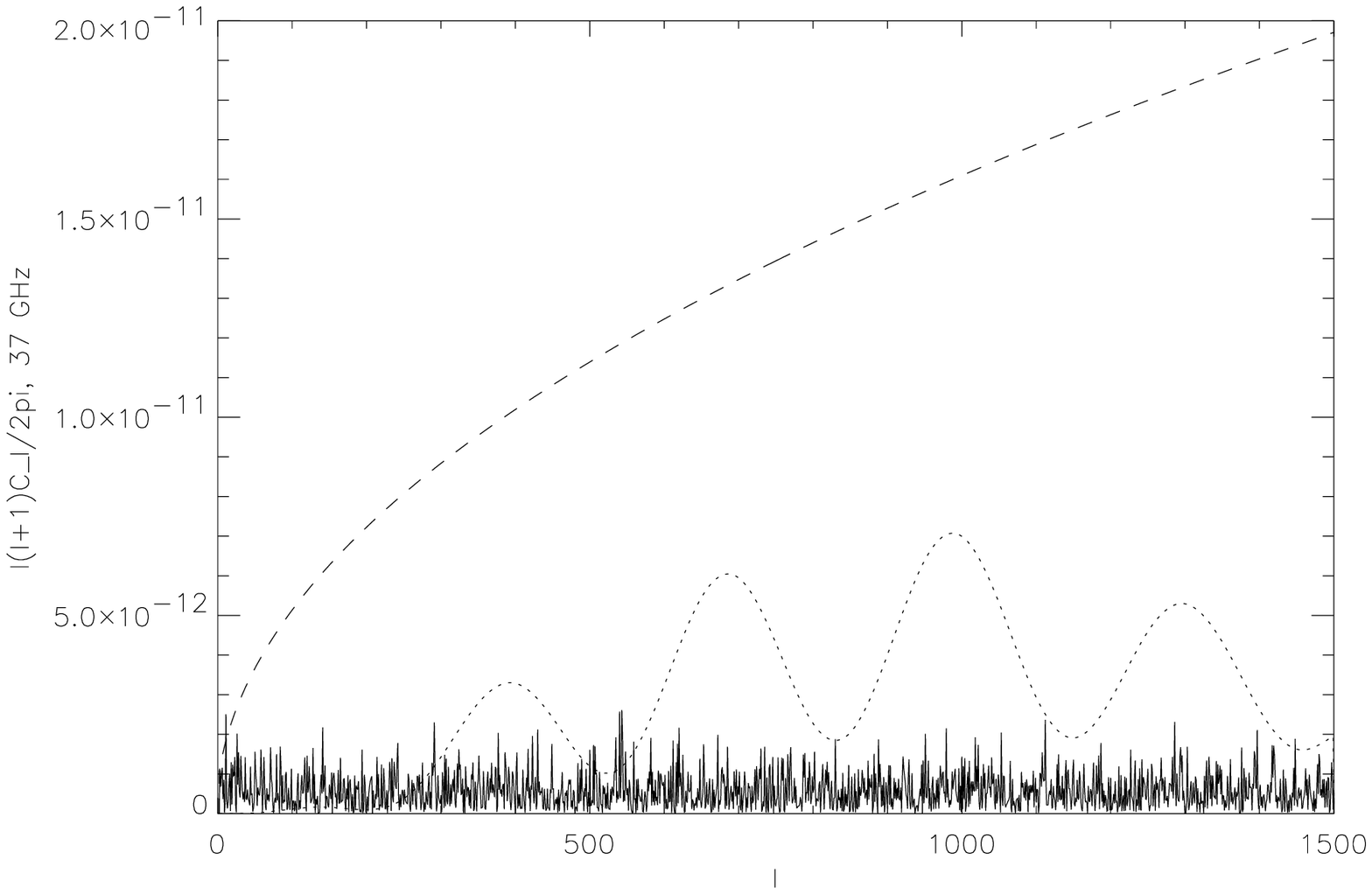,height=3.5in}}
\vspace{-12pt} \caption{\small Similar to figure \ref{fD}. Here we
plot simulated measurements with infinite angular smoothing
resolution (no smoothing on $l$) instead of the expectation value
of the estimator at 37 GHz for a synchrotron dominated sky. The
dotted line give both the estimator and the CMBP power spectrum
for a CMBP dominated sky. The dashed line is the power spectrum of
the galacic synchrotron when its expected contribution is maximum
(eq. (\ref{synEBB}) with $\beta = 2.9$, $\alpha = 1.5$ and $C_o =
2.6$) (see text for details)} \label{f10}
\end{figure}

\begin{figure}
\vspace{-15pt}
\centerline{\hspace{30pt}\epsfig{file=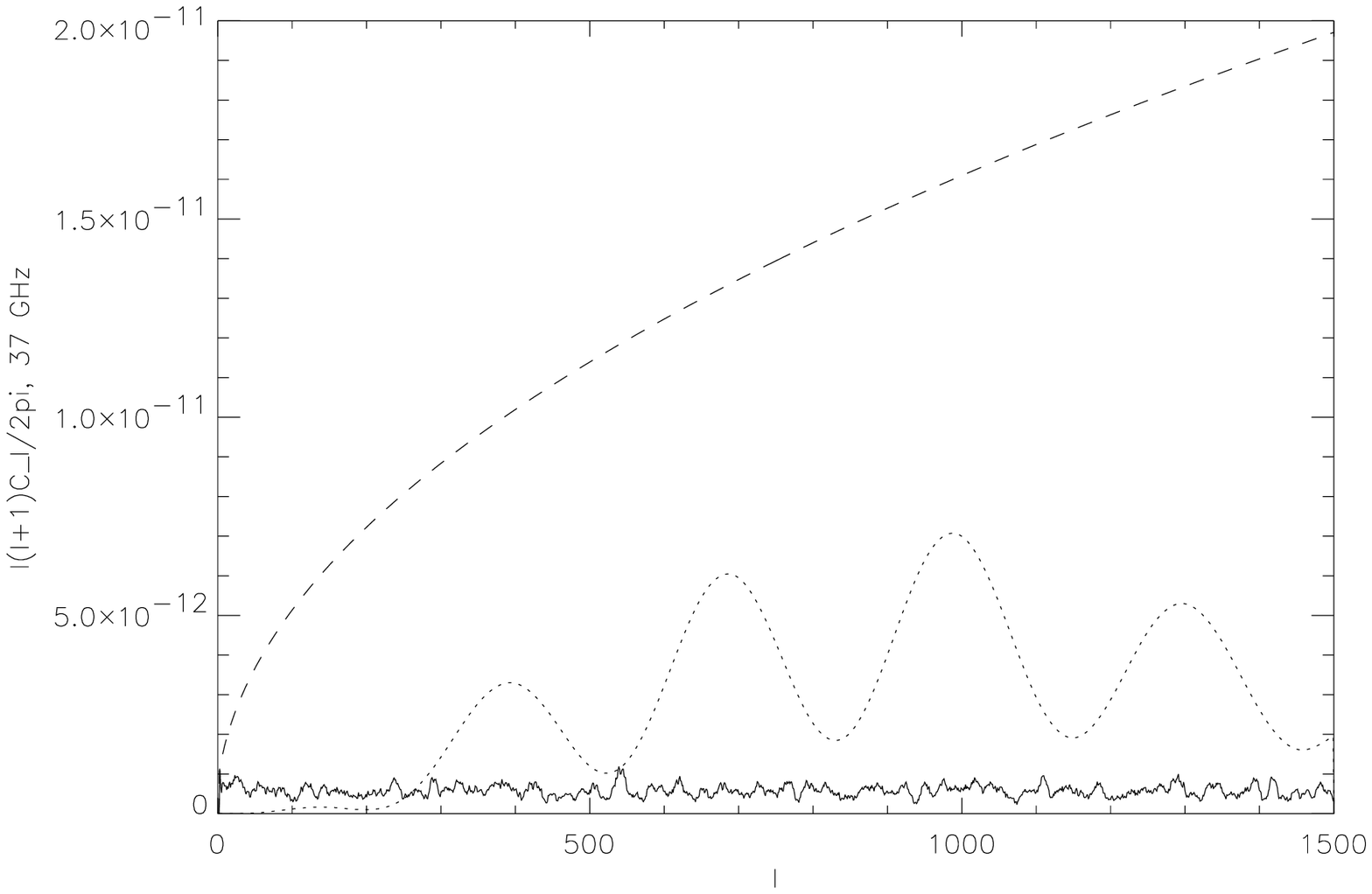,height=3.5in}}
\vspace{-12pt} \caption{\small Same as figure \ref{f10} with
finite angular resolution (smoothing on $\Delta l = 10$)}
\label{f11}
\end{figure}

\begin{figure}
\vspace{-15pt}
\centerline{\hspace{30pt}\epsfig{file=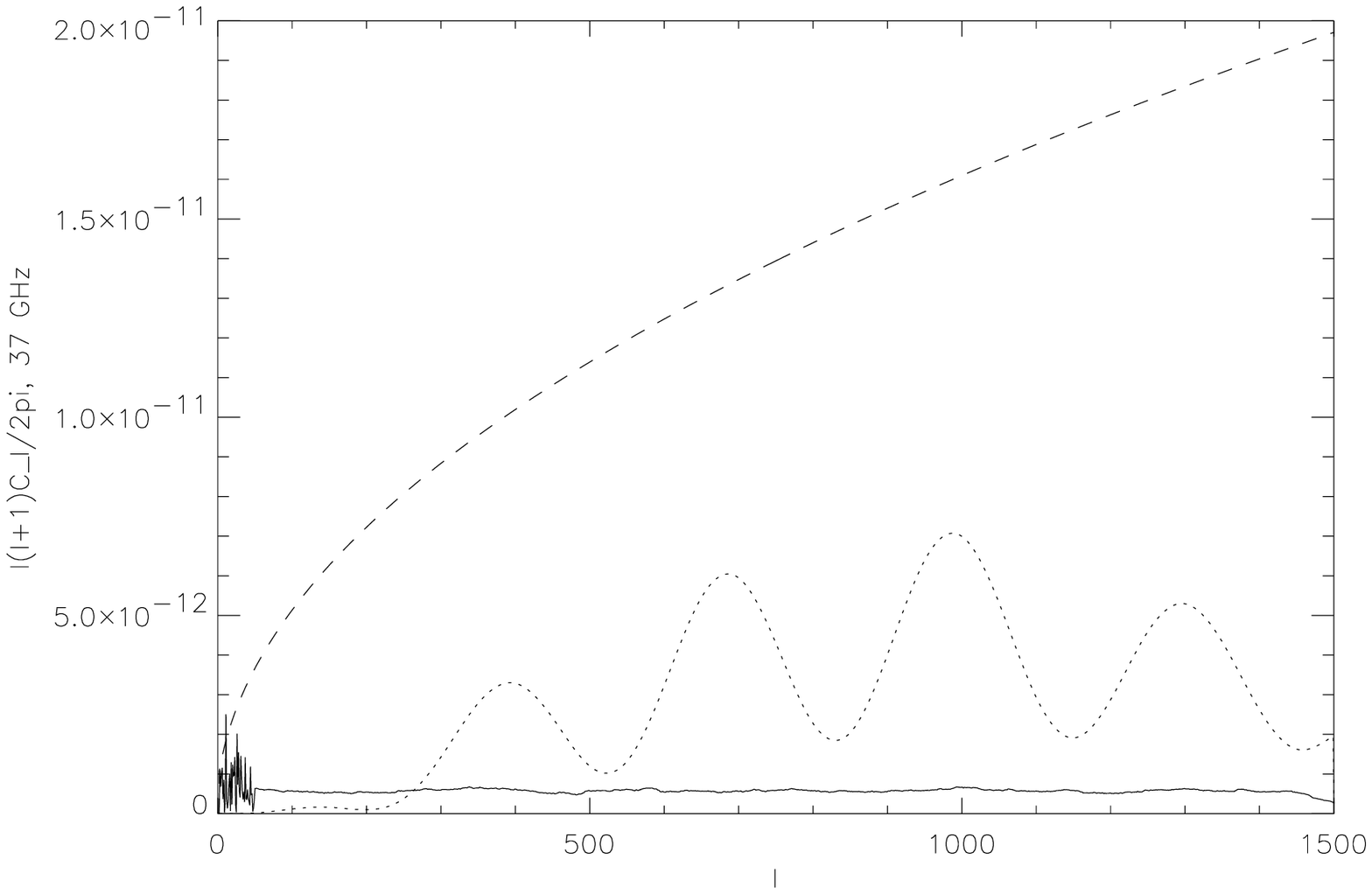,height=3.5in}}
\vspace{-12pt} \caption{\small Same as figure \ref{f10} with
finite angular resolution (smoothing on $\Delta l = 100$)}
\label{f12}
\end{figure}

Figure \ref{f10}, figure \ref{f11} and figure \ref{f12} are similar to figure
\ref{f10} and plot at 37 GHz, instead of the expectation value of $D^s_l$,
simulated measurements of it, for $\Delta l =1$ (figure \ref{f10}), $\Delta l
=10$ (figure \ref{f11}) and $\Delta l = 100$ (figure \ref{f12}),
respectively. Once again the CMBP power spectrum has been calculated  with
CMBFAST  assuming the same cosmological conditions we assumed above and the
synchrotron power spectrum comesfrom eq.(\ref{synEBB}) with $C_o=2.6,
\alpha=1.5, \beta=2.9$ (most pessimistic condition).

\begin{figure}
\vspace{-15pt}
\centerline{\hspace{30pt}\epsfig{file=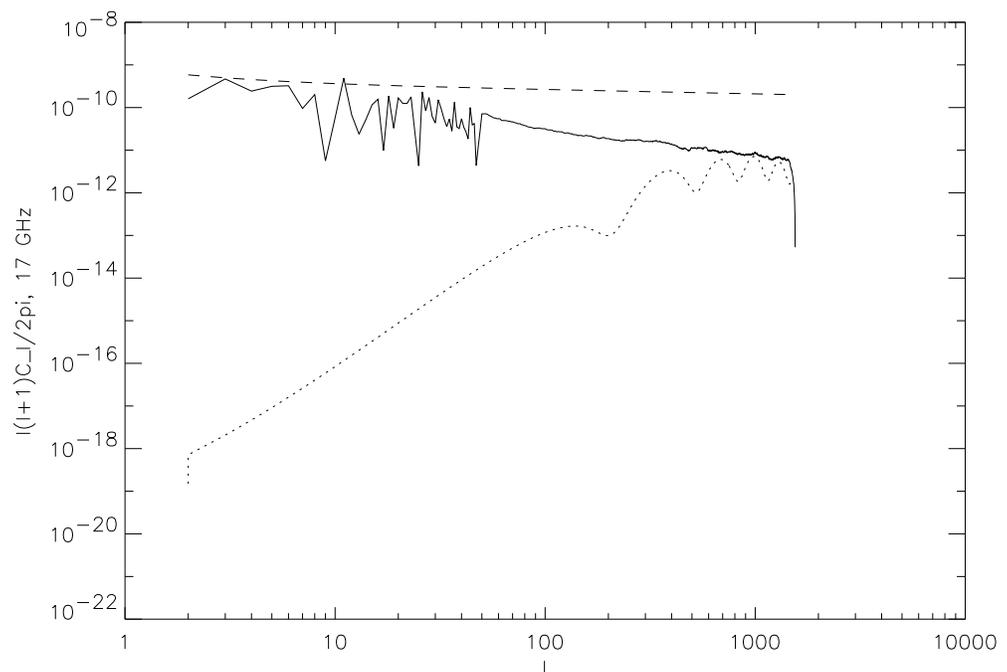,height=3.5in}}
\vspace{-12pt} \caption{\small Same as figure \ref{f12} at 17 GHz
using logarithmic scale on the vertical axis} \label{f13}
\end{figure}

Figure \ref{f13} is the same of figure \ref{f12} at a much lower frequency, 17
GHz. Here on the vertical axis we used logarithmic instead of linear scale,
which allows a better appreciation of the differences among the three curves.
The estimator power spectrum now almost touches the two highest peaks of the
CMBP spectrum. Therefore, 17 GHZ is probably the lowest frequency at which
one can hope to detect CMBP using $D_l$, in the most favorable conditions. In
the most pessimistic case ($C_0=2.6$ and $\alpha=1.5$) the corresponding
frequency is 25 GHz.

\section{Conclusions}
Observations of the CMB polarization are hampered by the presence of the
galactic polarized foreground. Only above $\sim 50 GHz$ the cosmic signal is
definitely above the galactic synchrotron and direct observations of CMBP are,
in principle, possible. Between $\sim 30$ GHz and $\sim 50$ GHz the level of
the polarized component of the synchrotron foreground is at least comparable
to the CMBP level, but the observational situation is still insufficient to
evaluate precisely its contribution to maps of the diffuse polarized
radiation measured by a telescope. Below $\sim 30$ GHz the galactic signal is
definitely dominant.
  So far a common approach for studies of CMBP was to make accurate
maps of the diffuse polarized radiation from a given region of sky at many
frequencies and disentangle the various contributions modelling their
frequency dependence and spatial distribution.

We have shown that an alternative way is to take advantage of the different
statistical properties of the spatial distribution of the main components of
the polarized diffuse radiation: CMBP and synchrotron (plus dust) galactic
foreground. By measuring the $E$ and $B$ modes of the polarized radiation we
can build an estimator which improves the background/foreground ratio by a
factor sufficient to allow firm recognition and extraction of the CMBP
contribution from single frequency maps at least down to 25 GHz (17 GHz in the
most favorable conditions) at angular scales $\leq 0.7^o$ ($l\geq 250$).

\section*{Acknowledgments}
We are indebted to S.Cortiglioni, E.Caretti, E.Vinjakin, J.Kaplan and
J. Delabrouille for helpful discussion. MVS
acknowledges the Osservatorio Astronomico di Capodimonte, INAF, for hospitality
during preparation of this paper.

\section*{Appendix: stochastic properties of
harmonics amplitudes}

 Here and overall in paper we suppose that $a_{lm}^s$
are complex random variables which satisfy the probability distribution law:
\begin{equation} \begin{array}{c}
p(a_{lm}^E) = \dfrac{1}{\pi E_l^2} exp\left( -\dfrac{|a_{lm}^E|^2}{
E^2_{lm}}\right) \\
~\\ p(a_{lm}^B) = \dfrac{1}{\pi B_l^2} exp\left( -\dfrac{|a_{lm}^B|^2}{
B^2_{lm}}\right)
\end{array} \end{equation}

\noindent
with variance
$\left< |a^E_{lm}|^2\right> =E^2_l$ and  $\left<|a^B_{lm}|^2\right> =B^2_l$.

They have all the
propertiers of gaussian variables (below we omitt indexes $E$ and $B$ in
first and second equations):

\begin{equation} \begin{array}{c}
\int\limits_{-\infty}^{\infty} p(a_{lm}) d^2 a_{lm} =1 \\
~\\ \int\limits^{\infty}_{-\infty}a_{lm} p(a_{lm}) d^2 a_{lm} =0
\end{array}
\label{gauss1}
\end{equation}

\begin{equation} \begin{array}{c}
\int\limits^{\infty}_{-\infty}|a_{lm}^E|^2 p(a_{lm}^E) d^2 a^E_{lm} =E^2_l \\
~\\ \int\limits^{\infty}_{-\infty}|a_{lm}^B|^2 p(a_{lm}^B) d^2 a^B_{lm} =B^2_l
\end{array}
\label{gauss2}
\end{equation}

Setting
\begin{equation}
\left< F\right> =\int\limits^{\infty}_{-\infty}F p(a_{lm}) d^2 a_{lm}
\end{equation}

\noindent
with current index $E$ or $B$, it immediately follows:
\begin{equation} \begin{array}{c}
\left<|a_{lm}^E|^4 \right> = \int\limits^{\infty}_{-\infty}|a_{lm}^E|^4 p(a_{lm}^E)
d^2 a^E_{lm} =2 E^4_l \\
~\\ \left<|a_{lm}^B|^4 \right> = \int\limits^{\infty}_{-\infty}|a_{lm}^B|^4
p(a_{lm}^B) d^2 a^B_{lm} =2 B^4_l
\end{array} \end{equation}

{}

\end{document}